\documentstyle[epsfig]{aipproc}
\newcommand{\Od}{{\cal O}}

\newcommand{\tr}{\mbox{tr}}

\newcommand{\fpi}{f_\pi}
\newcommand{\fpite}{f_\pi^t}
\newcommand{\fpisp}{f_\pi^s}
\newcommand{\ft}{f(t)}

\newcommand{\fdot}{\dot f(t)}
\newcommand{\fddot}{\ddot f(t)}

\newcommand{\intc}{\int_C dt \int d^3 \vec{x}}
\newcommand{\intcc}{\int_C d^4x}
\newcommand{\vxt}{(\vec{x},t)}

\newcommand{\be}{\begin{equation}}
\newcommand{\ee}{\end{equation}}
\newcommand{\ba}{\begin{eqnarray}}
\newcommand{\ea}{\end{eqnarray}}

\newcommand{\NP}[1]{{\it Nucl.\ Phys.\ }{\bf #1}}
\newcommand{\ZP}[1]{{\em Z.\ Phys.\ }{\bf #1}}

\newcommand{\PL}[1]{{\em Phys.\ Lett.\ }{\bf #1}}

\newcommand{\AN}[1]{{\em Ann. Phys. } (N.Y.) {\bf #1}}

\newcommand{\PR}[1]{{\em Phys.\ Rev.\ }{\bf #1}}

\newcommand{\IJmp}[1]{{\em Int.\ J.\ Mod.\ Phys.\ }{\bf #1}}

\def\IN{\relax{\rm I\kern-.18em N}}
\def\IR{\relax{\rm I\kern-.18em R}}
\def\ID{\relax{\rm I\kern-.18em D}}

\newcommand{\dnote}[1]{}

\begin{document}

\title{Nonequilibrium chiral perturbation theory and Disoriented
  Chiral Condensates
\thanks{Talk given in  ``Hadron Physics: Effective Theories of Low Energy
QCD'', Coimbra, Portugal.
}
}

\author{Angel G\'omez Nicola
\thanks{E-mail: 
gomez@eucmax.sim.ucm.es}
}
\address{ Departamento de F\'{\i}sica 
Te\'orica\\ Universidad Complutense, 28040, Madrid, Spain}

\maketitle


\begin{abstract}
We analyse the extension of Chiral
 Perturbation Theory to describe  a meson gas out of thermal equilibrium.
 For that purpose, we let the pion decay constant be a time-dependent
 function  and work within the
 Schwinger-Keldysh contour technique. A useful connection with curved 
space-time
 QFT  allows to consistently renormalise the model, introducing two
 new low-energy constants in the chiral limit. 
 We discuss the applicability of our approach within  
a Relativistic Heavy-Ion Collision
 environment. In particular, we investigate the  
 formation of Disoriented Chiral Condensate domains in this  model, 
via  the parametric resonance mechanism. 
\end{abstract}

\section*{Introduction and motivation}

The forthcoming 
experiments on  Relativistic Heavy Ion Collision (RHIC) at BNL and 
CERN will be able to test accurately  the dynamics 
 of the QCD plasma. After the collision, the plasma formed in the
 central rapidity region cools down via hydrodynamic expansion, and 
 nonequilibrium effects become important in that regime. Among them, 
  one of
 the most interesting suggestions has been the formation of the 
so called Disoriented
 Chiral Condensates (DCC). The DCC were 
proposed originally in  \cite{an89} 
as misaligned vacuum
 regions, where the chiral field points out in directions in isospin
 space different from that where the vacuum expectation value of the 
pion field vanishes. If such regions were formed, one could observe large
 clusters  of pions emitted coherently from the plasma as the pion
 field relaxes to the normal vacuum. This kind of behaviour is indeed
 observed in  Centauro and anti-Centauro events  in
 cosmic ray experiments \cite{centauro}. However, 
 one should point out  that a
 clear signal for DCC formation has not been observed yet in  the RHIC 
at Fermilab \cite{fermilab}, although it could well happen that the
DCC's are too small to be directly detected and one has to think of
alternative observables (see below).

On the other hand,
 after the hadronisation time, a proper description of the microscopic 
 meson dynamics makes it compulsory to use an effective low-energy 
 theory for QCD. In this context, the chiral symmetry 
plays a fundamental role. The effective theory must incorporate all
the QCD symmetries and the  chiral spontaneous symmetry breaking (SSB) 
pattern,
so that the Nambu-Goldstone bosons (NGB) are the lightest mesons
($\pi$, $K$, $\eta$). The light quark masses are then   
 introduced perturbatively. 
One possible choice is simply an 
 $O(N)$  model, with the standard classical SSB potential. 
  Its  fundamental fields  are 
 the $N-1$ pions and the $\sigma$, which acquires a nonzero 
 vacuum expectation value  $v$.  This is the  Linear
 Sigma Model (LSM) description.  However, one should bear  in mind
that the LSM becomes nonperturbative in the coupling constant at low
energies, so that  alternative perturbative expansions have to be used, 
such as  large $N$. 
 Besides, the LSM only  shares 
 the QCD chiral symmetry breaking 
 pattern  for $N=4$. 
An alternative approach is to construct an effective theory  as an 
 infinite sum of terms with increasing number of derivatives, 
only for the  NGB fields. 
 This is the description based in the  Nonlinear Sigma Model
 (NLSM), which is the lowest order action one can write down in this
 expansion. Higher order   corrections   come both from  NGB loops and 
 higher order lagrangians and  can be  renormalised 
 order by order in energies, yielding  finite predictions
 for  the meson observables. The unknown
 coefficients, which  encode all the information on the
 underlying theory,  absorb the loop infinities and 
 can be fitted to experiment. 
 This framework constitutes the so called chiral perturbation theory (ChPT) 
 \cite{we79,gale}. 
 The  perturbative expansion is carried out in terms of the ratio of 
 the $\Od(p)$ meson 
 energy scales of  the theory 
(masses, external momenta, temperature
 and so on) and 
  the  chiral scale $\Lambda_{\chi}\simeq$ 1 GeV 
(see \cite{dogoho92,meiss93,pich95,dogolope97} for a 
 review). 


Nonequilibrium effects such as the DCC's 
 have been investigated in the literature using $O(N)$ models with  initial
 thermal equilibrium conditions $\sigma (t=0)=0$ and $\pi^a (t=0)=0$.  In
this context, two different  scenarios for DCC formation have been
proposed: 
the first one takes place in the early stages of the plasma evolution. 
Roughly speaking, as the field rolls down along 
the potential, long wavelength modes grow exponentially and enhance
the formation of DCC's. There have been several approaches  in the
literature to implement this idea in the $O(N)$ model, like 
 classical simulations \cite{rawi93},  large $N$
\cite{bodeho95}, or analysis based on reasonable assumptions on the  
kinematics  \cite{cooper95,lamp96}. 
Typical DCC sizes within this approach are of the order of 2-3 fm,
containing  {$n_\pi\simeq$ 0.2 fm$^3$} pions, whereas  
the plasma cools down  in a  proper time of about  
$\tau\simeq$ 5-10 fm/c. As  commented above, 
these numbers yield too small DCC's to be 
observed directly. A second suggestion, which has been recently
proposed \cite{mm95,kaiser,hiro} is based on the parametric resonance
mechanism and inherits the idea from inflationary reheating
\cite{linde}. In this approach, the $\sigma$ field is very close to
the bottom of the potential but it is still {\em oscillating} around
it (it clearly overshoots the vacuum if the initial conditions 
 are imposed on the top) in the 
late stage of the plasma evolution. Those oscillations transfer energy 
to the pion modes, giving rise to exponentially growing pion solutions
for certain bands in momentum space, via parametric resonance. Recent
work within the LSM in this approach predicts rather larger DCC, of
sizes up to 5 fm \cite{kaiser}.  
More details about this mechanism will be given below.

In the present 
 work, we will explore the applicability of ChPT to describe the 
meson plasma out of thermal equilibrium. So far, this formalism has been
 applied only in  equilibrium, to
 study  the low $T$ ($T=\Od(p)$) meson gas
 and the chiral phase transition  \cite{gele89,boka96}.
 The key idea of our approach  is to make use of the
 derivative expansion consistently  defined  in ChPT in order to
 study the system not far from equilibrium. It is therefore best
 suited for the late stage evolution and has the additional 
  advantages typical
  of standard ChPT, i.e, it deals 
 only with NGB fields and is equally
 applicable to three flavours. We will show that a systematic  power
 counting can be established  in this case and, furthermore, that 
 the renormalisation program can be consistently 
implemented. Details can be
 found in \cite{agg99}. In addition, in the last section 
we will explore the possibility of describing DCC formation 
within ChPT, via parametric
 resonance.

\section*{The model and chiral power counting}
 Our starting point is the nonlinear sigma model (NLSM) where we
 let the pion decay constant-- the only relevant parameter to the lowest
 order in derivatives-- be time dependent. In the context of a RHIC, such time
 dependence can
 be thought of as proper time evolution within the so called Bjorken
 initial conditions \cite{bjo83}, where observables depend only on
 proper time and not on rapidity. This picture is consistent with the
 experimental observations. 
 We take the
  initial time $t=0$, having in mind that it  would correspond
   to a proper time $\tau_0\simeq$ 1-2 fm/c, a typical
 hadronisation time. 
  Thus, we will consider the following NLSM action
\begin{equation}
S[U]=\intc \ \frac{f^2(t)}{4} \ \tr \ \partial_\mu U^\dagger\vxt
\partial^\mu U\vxt
\label{nlsmne}
\end{equation}

Here, $C$ is the Schwinger-Keldysh contour  (see \cite{agg99} for
details), which parametrises the nonequilibrium path integral where
we are considering  thermal equilibrium for $t\leq 0$ at a temperature
$T_i=\beta_i^{-1}$,  as the initial condition. 
Note that the action (\ref{nlsmne}) is chiral invariant
($U\rightarrow LUR^\dagger$) by construction, which will play an important 
role in what follows.
 As a first approximation, we will be
 interested only in the  strict chiral limit for two
light flavours,  i.e,  massless pions. Therefore, 
 we are not including any explicit symmetry-breaking term in the action. 
 Thus, $f(t\leq 0)=f\simeq$ 93 MeV
 to leading order ($f\neq \fpi$ to higher orders) and 
 for $t>0$ the system departs from equilibrium.
 Note that, since we choose that departure to be instantaneous,
 $\ft$ cannot be analytical at $t=0$. This is just an artifact of the
 approximation and should not have any effect on the long-time
 behaviour.  Finally, as customary, 
$U(x)$ is  parametrised in terms of pion fields $\pi^a$ as:
\ba
U\vxt=\frac{1}{f(t)}\left\{\left[f^2(t)-\pi^2 \right]^{1/2} I
+i\tau_a\pi^a\right\}
\label{upar}
\ea
and  $\pi^a(t_i-i\beta_i)=\pi^a(t_i)$ is the equilibrium boundary
condition, with $t_i<0$. 


The new ingredient we need to incorporate in the power counting in order
 to be consistent with ChPT is then
\begin{equation}
\frac{\fdot}{f^2 (t)}\simeq \Od\left(\frac{p}{\Lambda_\chi}\right), \qquad
\frac{\fddot}{f^3(t)}, \frac{[\fdot]^2}{f^4(t)}
\simeq\Od \left(\frac{p^2}{\Lambda_\chi^2}\right),
\label{chipoco}
\end{equation}
and so on.  Obviously, our results will depend upon the  choice of
$\ft$. One can think
of $\ft$ as an external source, to which we wish to obtain the
nonequilibrium response of the system. Alternatively, this model 
 can be thought of to lowest order as the  LSM with the time-dependent
constraint $\sigma^2+\pi^2=f^2(t)$. We shall discuss 
 below a reasonable assumption  for 
 $\ft$ in connection with  DCC
 formation. Meanwhile, we shall keep $\ft$ arbitrary.

To lowest order in the pion fields, the above NLSM action
 can be written as 
\begin{eqnarray}
S_0[\pi]&=&- \frac{1}{2}\intcc\pi^a\vxt\left[  \Box + 
m^2 (t)\right] 
\pi^a\vxt 
\label{act2pi}
\end{eqnarray}
where $\intcc=\intc$ and 
$m^2(t)=-\fddot/\ft$. That is, the model accommodates a
time-dependent pion mass term, without breaking explicitly the chiral
symmetry. This effect is the same as switching on an external curved
space-time background, as we will see in the next section.

\section*{Renormalisation and curved space-time}
Once we have defined our nonequilibrium power counting, we can apply
 ChPT to calculate the time evolution of the observables.
 In doing so, we must pay special attention to renormalisation.
The fact that there is a time-dependent mass term indicates that there 
can be new time-dependent infinities in the chiral loops. However, we
are in the chiral limit, so we are not allowed to introduce the 
usual $\Od(p^4)$ mass and wave function counterterms breaking the
chiral symmetry \cite{gale}. In other
words, we should be able to construct the most general fourth order
action, which in particular should include new terms (and hence new
low-energy constants) to cancel those extra  divergences, preserving
exactly the chiral symmetry. 

In order to find this $\Od(p^4)$ lagrangian, 
 we will make use of a very fruitful analogy: the action
 (\ref{nlsmne}) is equivalent to formulate the
   NLSM on a curved space-time background
 corresponding to a spatially flat Robertson-Walker metric, with
 scale factor  $a(t)=f(t)/f(0^+)$ (see \cite{agg99} for details). 
 Note that in this language,
 $m^2(t)$ in (\ref{act2pi}) represents  the minimal coupling with the RW
metric preserving chiral
 invariance.

  Therefore, we can construct the
 $\Od(p^4)$ action as:
 \begin{eqnarray}
S_4 [U,g,R]=\intcc \sqrt{-g}\left[
{\cal L}_4[U,g]+
(L_{11}R 
g^{\mu\nu}+L_{12}R^{\mu\nu}) \tr 
\partial_\mu U^\dagger 
\partial_\nu U\right]
\end{eqnarray}
where  $g$
 is the metric determinant, 
${\cal L}_4[U,g]$ stands for  the standard (equilibrium) lagrangian
\cite{gale}
 with indices raised and lowered with the $g^{\mu\nu}$ metric and 
the rest are new $\Od(p^4)$ invariant 
couplings with  the scalar curvature $R(x)$ and the Ricci
 tensor $R_{\mu\nu} (x)$ in the chiral limit. 
These are the new terms we need, where
 $L_{11}$ and $L_{12}$ are the new coupling constants. 
In fact, this problem has been already considered   in
 \cite{dole91} in order to study the  
energy-momentum tensor of QCD at low energies. In that work it has
been 
found that $L_{11}$ is renormalised in dimensional regularisation,
whereas $L_{12}$ is already finite. Their numerical  values can be
obtained from the experimental information on  the QCD energy-momentum form
factors. They yield  $L_{12}\simeq -2.7\times 10^{-3}$ and
$L_{11}^r(\mu=1 GeV)\simeq 1.4\times 10^{-3}$ where $\mu$ is the
renormalisation scale. In our case, with our RW metric we get to
$\Od(\pi^2)$, 
\begin{eqnarray}
S_4[\pi,g]= - \frac{1}{2}\intcc\pi^a\left[f_1 (t)\partial_t^2-f_2 (t)\nabla^2
 + m_1^2 (t)\right]\pi^a +\Od(\pi^4)
\label{lag4}
\end{eqnarray}
with
\begin{eqnarray}
f_1(t)&=&12\left[\left(2L_{11}+L_{12}\right)\frac{\fddot}{f^3(t)}
-L_{12}\frac{[\fdot]^2}{f^4 (t)}\right]
\nonumber\\
f_2(t)&=&4\left[\left(6L_{11}+L_{12}\right)\frac{\fddot}{f^3(t)}
+L_{12}\frac{[\fdot]^2}{f^4 (t)}\right]
\nonumber\\
m_1^2 (t)&=&-\left[\frac{f_1(t)\fddot+\dot f_1 (t)\fdot}{\ft}+
\frac{1}{2}\ddot f_1 (t)\right]
\label{f12}
\end{eqnarray}

The above lagrangian should take care of the nonequilibrium
infinities we might find in the pion two-point function. We will see
below that this is indeed the case.

\section*{The pion decay functions $\fpi (t)$}

The first observable one might think of calculating in ChPT is the  
pion decay constant to one loop. In the nonequilibrium model, it will
become a time-dependent function $\fpi (t)$. One should point out that 
the definition of $\fpi$ is subtle even in thermal equilibrium
\cite{boka96,kash94}. In addition,  one has in general $\fpi^s(T)\neq
\fpi^t(T)$
 corresponding to the  axial current spatial and temporal components  
 and due to the loss of Lorentz covariance in the
 thermal bath \cite{pity96}. 
We refer to \cite{agg99} for details on how to define
 properly $\fpi (t)$ out of equilibrium. Once this has been done, one has
 to consider the one loop diagrams for the pion two-point function 
 coming from 
 (\ref{nlsmne}) plus the tree level ones from (\ref{lag4}). The final
 result up to $\Od(p^4)$ reads \cite{agg99}
\ba
\left[\fpisp (t)\right]^2&=&f^2 (t)\left[1+2f_2(t)-f_1(t)\right]-2i G_0(t)
\label{fpispnlo}\\
\left[\fpite (t)\right]^2&=&f^2 (t)\left[1+f_2(t)\right]-2i G_0(t)
\label{fpitenlo}
\ea
for $t>0$, with $f_{1,2}(t)$ in (\ref{f12}) and  $G_0 (t)$ is nothing but the
 equal-time pion two-point function  $G_0(t)=G_0(x,x)$ with
 $G_0(x,y)$ the solution of the differential equation
\begin{equation}
\left\{\Box_x + m^2(x^0)\right\} G_0 (x,y)=-\delta_C (x^0-y^0)
\delta^{(3)} (\vec{x}-\vec{y})
\label{loprop}
\end{equation}
with   KMS equilibrium  conditions $G^>_0
(\vec{x},t_i-i\beta_i;y)=G^<_0 (\vec{x},t_i;y)$, $G_0^{>}$ and
$G_0^{<}$ being advanced and retarded correlation functions. 
Clearly, this
equation cannot be solved analytically for an arbitrary $\ft$, but it
can  be managed numerically. Therefore, one must remember that
$G_0 (t)$ depends implicitly on $\ft$ through (\ref{loprop}).

As a consistency check, 
the results (\ref{fpispnlo})-(\ref{fpitenlo}) reproduce the
equilibrium result \cite{gale87}
 when we switch off  the time derivatives of
$\ft$:
\begin{eqnarray}
\left[\fpi^{s} (T)\right]^2=\left[\fpi^{t} (T)\right]^2
=f^2\left(1-\frac{T^2}{6 \fpi^2}\right)
\label{equi}
\end{eqnarray}

An interesting
 consequence of our result is that  $\fpisp (t)\neq \fpite (t)$ to
 one-loop, unlike the equilibrium case. In addition,
 from (\ref{fpispnlo})-(\ref{fpitenlo}) and (\ref{f12}) we see that the
 difference $[\fpisp (t)]^2-[\fpite (t)]^2$ is finite, so that 
  $\fpi^{s}(t)$ and  $\fpi^{t}(t)$ can be renormalised 
at the same time, which is another consistency check.
  We remark that $G_0(t)$ contains in general UV divergences,
 to be absorbed by $f_1(t)$
 and $f_2(t)$  in the renormalisation of $L_{11}$. An explicit check
 of this renormalisation procedure 
 will follow in the next section.

\section*{Disoriented Chiral Condensates in ChPT}

In this section we will consider a particular choice of $\ft$ and
apply our previous results. Our motivation is the 
possibility of generating DCC-like structures in this context. We
shall sketch some of our preliminary results here, while  
details of the calculation and further work 
will be postponed to a forthcoming paper.

As we
have discussed above, our approach is meant to be useful in a stage of 
the plasma evolution where the departure from equilibrium is of the
same order as the meson energies. Hence, we should be able to obtain
similar results as the analysis performed in the LSM in the parametric 
resonance regime \cite{mm95,kaiser,hiro}, where the rolling down of
the $\sigma$ field is in its late oscillatory period. This is the same 
behaviour of the inflaton field in reheating  \cite{linde}. 
 One then allows for a time-dependent classical
background $\sigma (t)$ in the LSM, splitting the  field as
\be
\sigma (\vec{x},t)=\sigma (t)+\delta\sigma(\vec{x},t)
\ee
where $\delta\sigma$ is the
quantum fluctuation. As a first approximation, one can  neglect
 the pion fluctuations  
$\langle \pi^2 \rangle \ll v^2$ \cite{mm95,hiro} 
and solve the equation of motion, which yields
just  $\sigma(t)=\sigma_0 \cos m_\sigma t$. Here,  $\sigma_0$
is the initial field amplitude, which in this approximation is
a small quantity. Even though, one can still produce 
exponentially growing pion fields (DCC) which in the end will be
responsible for the damping of the oscillations as the field relaxes
to equilibrium. One should bear in mind that neglecting $\langle \pi^2
\rangle$ to lowest order is a rather crude approximation, as pointed out in
\cite{kaiser}, which is clearly not valid for large times when the
pion correlator grows significantly. Nonetheless, we will carry on
with this simple case, just to understand qualitatively 
how ChPT can also account for the description of DCC's. A better
approximation would be  to solve the coupled equations for the $\sigma$ and
$\pi$ fields, which yields the solution for $\sigma(t)$ in terms of
elliptic functions  \cite{kaiser}. 
Therefore, in this simple picture, we take  our $\ft$ of the same
form as the lowest order $\sigma(t)$ in the LSM, i.e, 
\be
f(t)=f\left[1-\frac{q}{2}\left( \cos Mt -1\right)\right]
\ee

\begin{figure}[b!]
\centerline{\epsfig{file=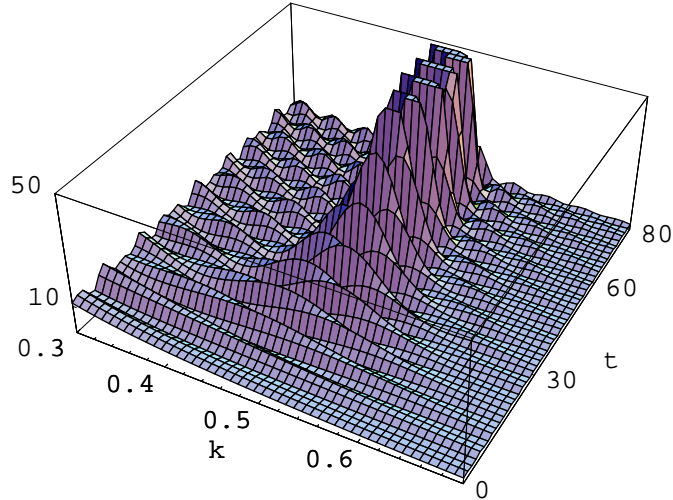,width=11cm}}
\vspace{10pt}
\caption{ $iMG_0 (k,t)$ for $T_i=M$ and $q=0.1$. Spatial 
  momentum and time are measured in units of $M$ and $M^{-1}$
  respectively. The instability band 
 for this case lies roughly between $0.4M<k<0.6M$.}
\label{fig1}
\end{figure}

Here, $q$ is a small parameter, playing  the role of $\sigma_0$ in
the LSM. Notice that our nonequilibrium chiral power counting demands
$q M^2=\Od (p^2)$ and so on. Thus, for definiteness, we will take
$q=\Od(p^2/\Lambda_\chi^2)$, so that the $\Od(p^4)$ corrections remain 
under control (see below), and  $M$  arbitrary. In the end, we will 
  discuss  how the results are affected by  $T_i$, $q$ and
$M$. Therefore, we have $m^2 (t)=-(qM^2/2)\cos Mt(1+\Od(q))$, so that  
the differential equation (\ref{loprop}) becomes to leading order
\be
\left[\frac{d^2}{d t^2} +  \frac{4 k^2}{M^2}-2q\cos M t\right]
 G_0^>(k,t,t')=0
\label{mathieu}
\ee
where we have Fourier transformed in the spatial coordinates only
($k^2=\vert\vec{k}\vert^2$). The above equation is nothing but the
Mathieu equation, which has several well-known interesting properties
\cite{mac,abramo}. Among them, it admits unstable solutions
exponentially growing in time, for certain values of
$4k^2/M^2$. This is the simplest version of the parametric resonance 
mechanism. 
 In particular, the instabilities develop in bands in $k$, centered at 
 $k_n=nM/2$,  of width $\Delta k_n=\Od(q^n)$. Hence, in the
 approximation we are working, only the first band is relevant, i.e,
 unstable solutions only exist for $M/2-\Delta k_1<k<M/2+\Delta
 k_1$. This is known in the Cosmology literature as the narrow
 resonance approximation \cite{linde}.  
A typical unstable solution $G_0 (k,t)$
has been plotted in Figure \ref{fig1} for a particular choice of the
parameters in the first band. 
The solutions typically oscillate with an exponentially
growing amplitude inside the unstable region. 
Therefore, we see that our ChPT approach allows for DCC-type
configurations.

Next, we will apply our results for $\fpi (t)$ to this
particular case. The equal-time correlation function 
$G_0 (t)=\int d^3 k G_0(k,t,t)$ turns out to be UV divergent, as expected. 
After standard  manipulations in  dimensional regularisation
($d=4-\epsilon$) 
one can cast the  divergent part for $t>0$ as
\be
i G_0^{div}
(t)=-\frac{qM^2}{16\pi^2}\cos Mt \left(\frac{1}{\epsilon}+
\frac{1}{2}\log\frac{\mu^2}{M^2}\right)
\ee
which is  an example of
the new time-dependent divergences we were talking about in previous
sections. In fact, we see that it 
has exactly the same form as $\ddot f(t)/f (t)$. Furthermore, replacing
in (\ref{fpispnlo})-(\ref{fpitenlo}), we find that 
the result is rendered {\em finite}
and {\em scale independent}  
with the same renormalisation of $L_{11}$ derived in \cite{dole91}.

\begin{figure}[t!]
\hspace*{.5cm}
\centerline{\hbox{\epsfig{figure=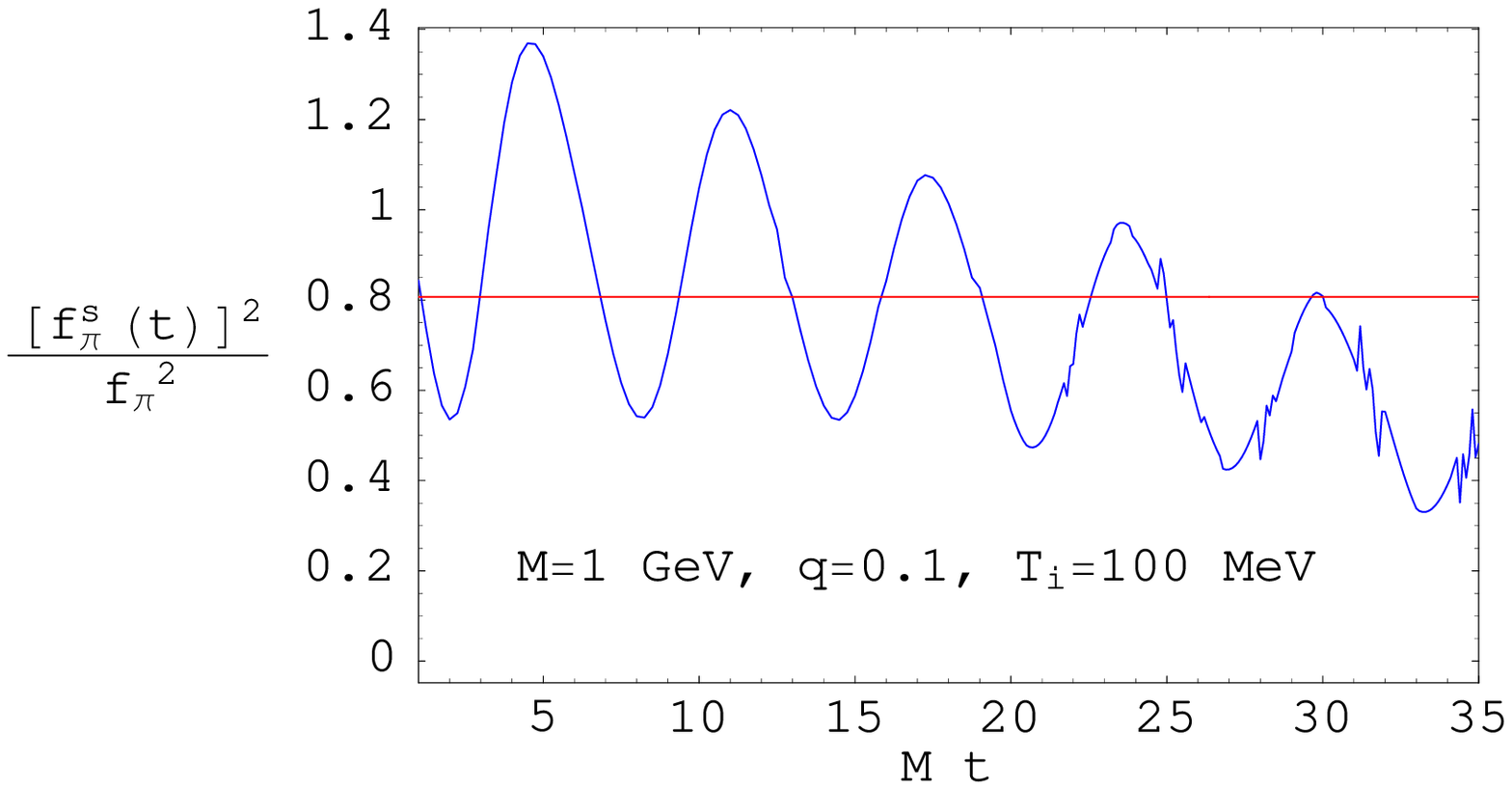,width=9cm}
\epsfig{figure=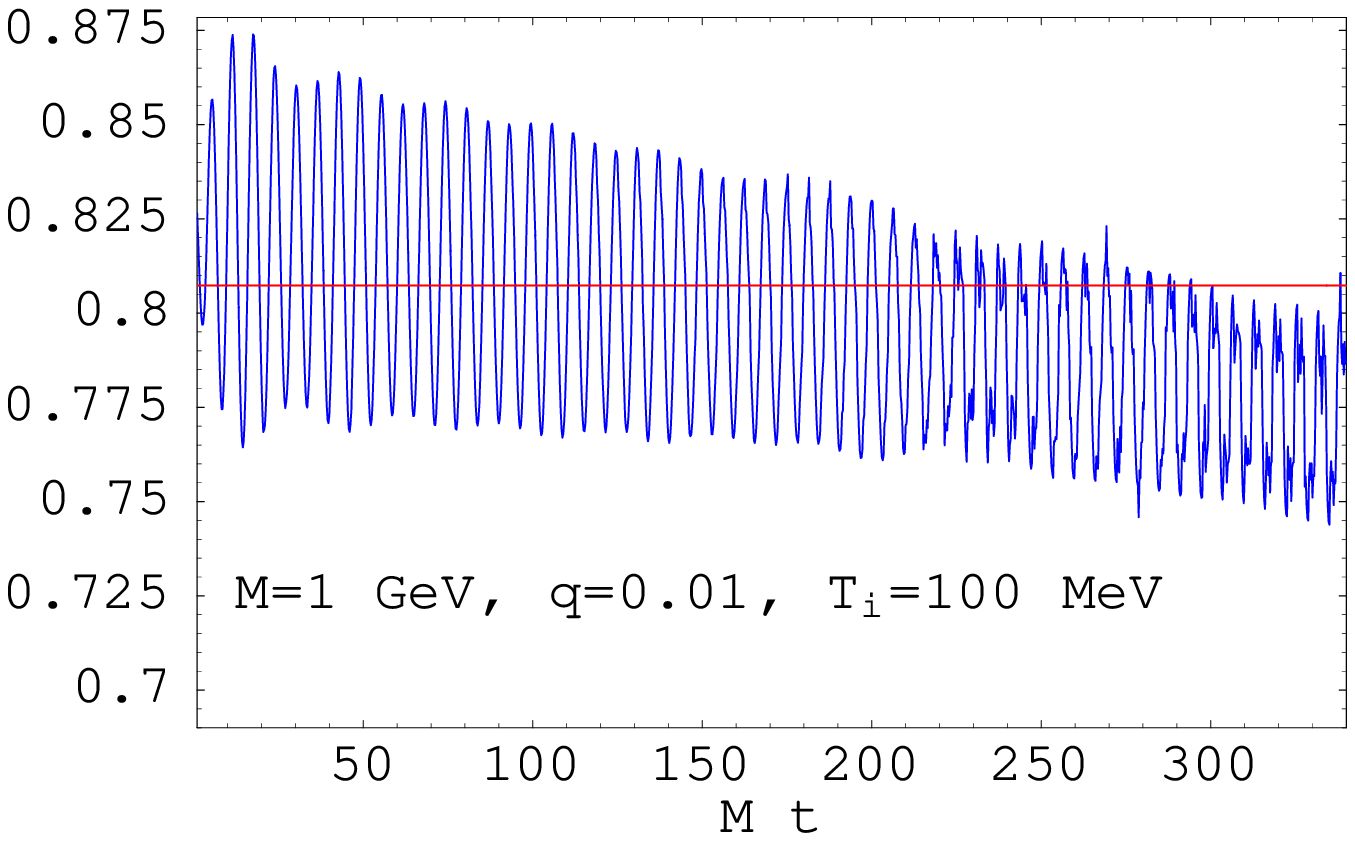,width=9cm}}}
\hspace*{-.4cm}
\hspace*{.5cm}
\centerline{\hbox{\epsfig{figure=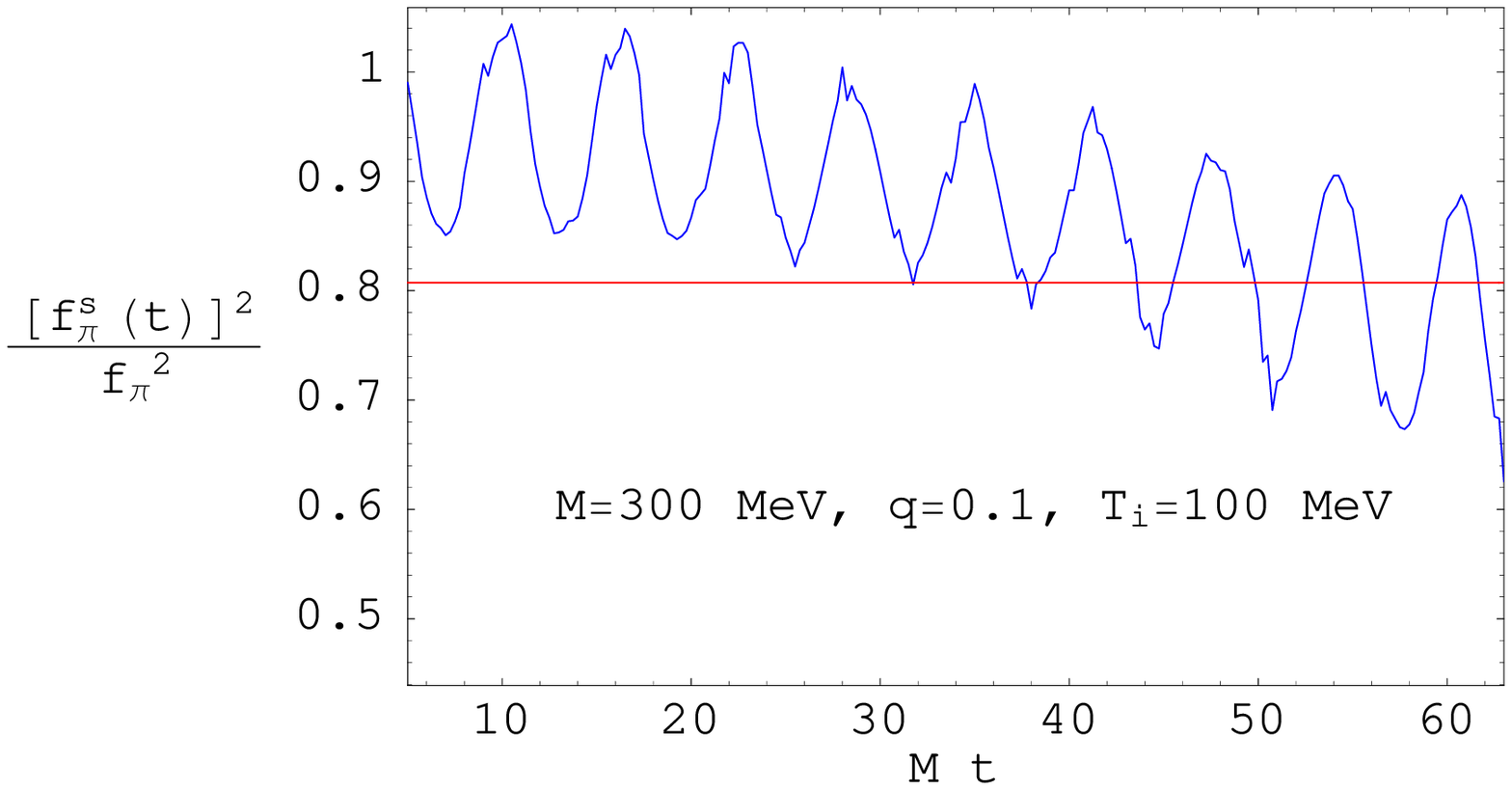,width=9cm}\hspace*{-.8cm}
\epsfig{figure=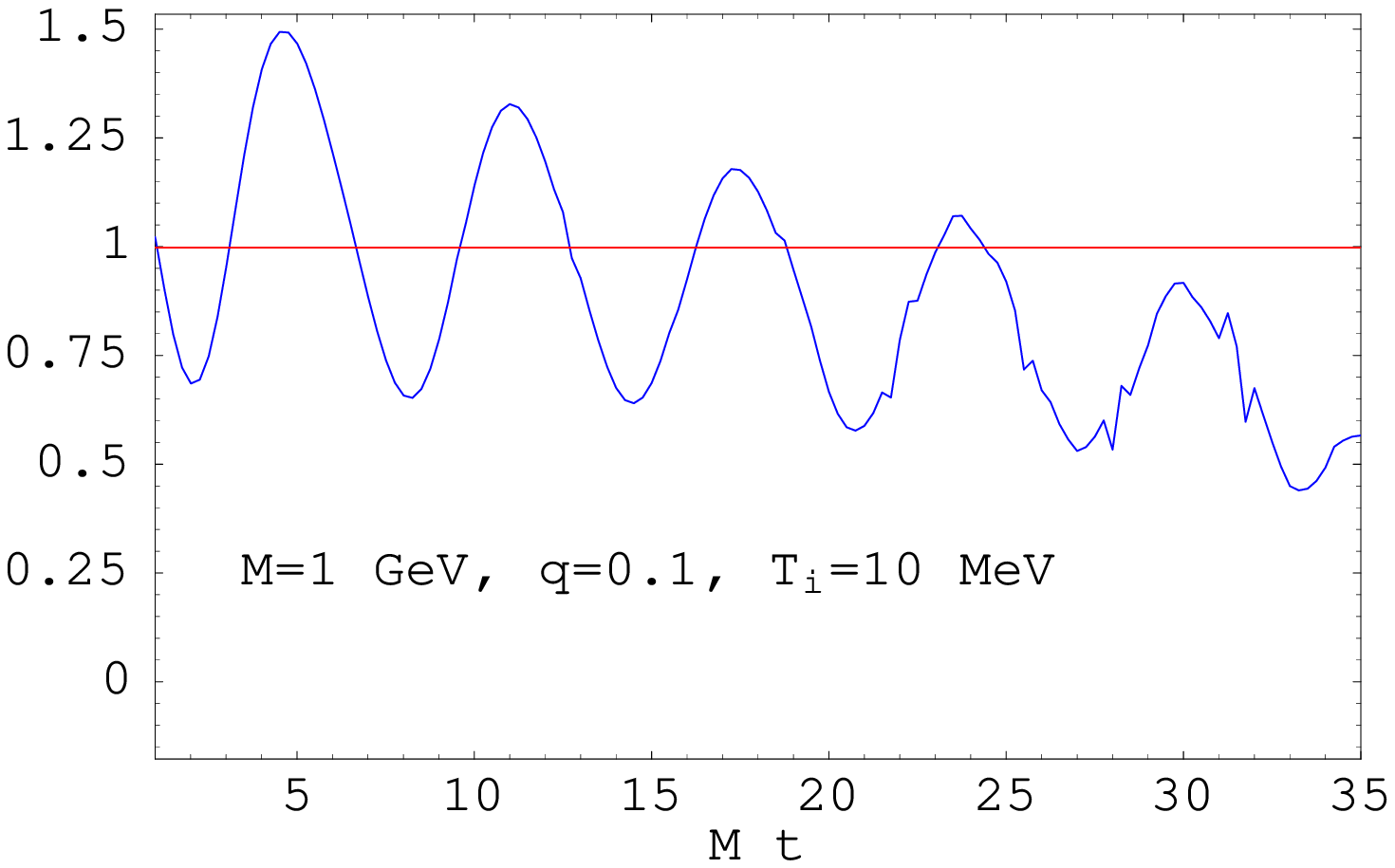,width=9cm}}}
\caption{Results for $\fpi^s (t)$ in the parametric resonance regime. 
The horizontal line is the
  value of $\fpi$ at the initial temperature $T_i$}.
\label{fig2}
\end{figure}

The final results for $\fpi^s (t)$ are plotted in Figure \ref{fig2} for
different choices of the parameters. 
We  clearly observe  the damping effect on the amplitude due to
the unstable solutions at long times. In other words, the DCC's
accelerate thermalisation. We also observe that this  mechanism
becomes less efficient for smaller $q$ and $M$. Typically, the
unstable corrections to the amplitude of $\fpi (t)$ are proportional
to $(qM^2/4\pi\fpi^2)\exp (qMt)$. 
On the other hand, this effect  seems to be rather insensitive
to the initial temperature and thus we expect to catch all the
important qualitative behaviour concerning the DCC's, 
 regardless of the initial conditions.  It should be pointed out 
that the curves have been cut off at the times where the one-loop
contribution becomes  of the same size as the tree level one. From
that point onwards, the exponentially growing correlator dominates,
yielding unphysical results. As
commented above, we do not expect our simple cosine shape for $\ft$ to 
be valid for all times, since it is derived neglecting the pion
correlator. This final time $t_f$  roughly defines 
the applicability range
of our results.  We expect that this range is enough to account for
all the plasma time evolution of a realistic RHIC. For instance, for $M$= 1
GeV and $q=0.1$, we get $t_f\simeq 35M^{-1}\simeq$ 8 fm/c.  
This is exactly the same  as extrapolating the
equilibrium result (\ref{equi}) to predict the critical temperature
at $T= T_c\simeq 6\fpi^2$, where 
all the higher order corrections become of the same
order. Nonetheless, that formula predicts the right behaviour of
$\fpi (T)$ as it approaches the transition. In the same way, our
results reproduce the expected qualitative behaviour as we extrapolate 
them up to times $t\simeq t_f$. 
Therefore,  $\fpi (t)$ can be regarded as an alternative observable (it
is the residue of an axial-axial correlator and it can be measured in
semileptonic decays) to test the size of
DCC-like configurations in the late stage of the plasma expansion.

\section*{Conclusions and Outlook}

We have reviewed recent work on the extension of 
 ChPT to a nonequilibrium situation. The NLSM with a time dependent
 pion decay constant provides  a nonequilibrium effective
 model with a well-defined perturbative expansion and power counting 
near equilibrium. 
The analogy of this model with  curved space-time QFT allows
 to consistently construct higher order lagrangians and 
implement renormalisation.  As a first application, we have
obtained the  renormalised one-loop  $\fpi (t)$. 

We have also shown how this
model can be applied to describe DCC-like structures in the late stage 
of the expansion of a hot plasma formed after a  RHIC. Work in
progress includes a more realistic study of the parametric resonances, 
including consistently the pion correlations in $\ft$ and calculating 
 the correlation length and the number of pions. 
One can also think of  including   pion
  masses, extending the results to three flavours, using large
  $N$ methods and cosmological applications 
as other interesting aspects to be investigated  in this  context.

\section*{Acknowledgments}
 I wish to thank the organisers of the  ``Hadron Physics'' conference
 and the Theory group in Coimbra for their kind help and hospitality. 
Financial
 support from CICYT, Spain,  project AEN97-1693, is  also acknowledged.

\end{document}